\documentclass[12pt]{article}
\usepackage{graphicx}
\newcommand{\odra}{O.~Dragoun}
\newcommand{\rys}{M.~Ry\v sav\'y}
\begin{document}
\renewcommand{\refname}{\large\bf References}
\newcommand{\bfl}{\begin{flushleft}}
\newcommand{\efl}{\end{flushleft}}
\newcommand{\bfr}{\begin{flushright}}
\newcommand{\efr}{\end{flushright}}
\newcommand{\beq}{\begin{equation}}
\newcommand{\eeq}{\end{equation}}
\newcommand{\bec}{\begin{center}}
\newcommand{\eec}{\end{center}}
\newcommand{\bit}{\begin{itemize}}
\newcommand{\eit}{\end{itemize}}
\newcommand{\plm}{$\pm$}
\newcommand{\bet}{$\beta$}
\newcommand{\ria}{$\rightarrow$}
\newcommand{\asi}{$\sim$}
\newcommand{\krat}{$\times$}
\def\up#1{$^{#1}$}
\def\dn#1{$_{#1}$}
\def\elem#1#2#3{$_{\ #1}^{#2}$#3}
\def\sct#1{\section{\large#1}}
\def\subsct#1{\subsection{\large#1}}
%
\def\cjp#1#2#3{{\it Czech. J. Phys.} {\bf #1}(#2)#3}
\def\np#1#2#3{{\it Nucl. Phys.} {\bf #1}(#2)#3}
\def\pl#1#2#3{{\it Phys. Lett.} {\bf #1}(#2)#3}
\def\pr#1#2#3{{\it Phys. Rev.} {\bf #1}(#2)#3}
\def\jpg#1#2#3{{\it J. Phys. G: Part. Nucl. Phys.} {\bf #1}(#2)#3}
\def\jpb#1#2#3{{\it J. Phys. B At. Mol. Opt. Phys.} {\bf #1}(#2)#3}
\def\zp#1#2#3{{\it Z. Physik} {\bf #1}(#2)#3}
\def\zetf#1#2#3{{\it Zh. Eksp. Teor. Fiz.} {\bf #1}(#2)#3}
\def\pc{private communication}
\def\ip{in preparation}
\bfr \it Report UJF-THE-01/2013 \efr
\bec
{\bf \Large Which one of the two atomic potential is better -- the newest or the ``ancient'' one?} \\ \ \\
\rys\footnote{E-mail: rysavy@ujf.cas.cz} \\
Nuclear Physics Institute, Acad. Sci. Czech Republic,
     CZ--250 68 \v Re\v{z} near Prague, Czech Republic
\eec
\begin{abstract}
Recently, there appeared the tables of atomic potential by Maldonado et al.
We show that this potential, when used to calculate some quantities which
can be compared with experiment, gives worse results than the more than 40
 years old tables by Lu et al.
\end{abstract}
\sct{Introduction}
  \label{s:intro}
In principle,  the atom is a very simple system. The nucleus in the center and
from one (for hydrogen) to \asi 100 (for transuranic elements) electrons
``orbiting'' around. However, except the hydrogen atom -- i.e. system of one
 proton and one electron -- it cannot be exactly described. It is well known
 that the problem of three or more bodies is not analytically solvable in
  classical mechanics, not to mention the quantum one.

Therefore a concept of atomic potential was introduced. Such potential is
an environment in which the electrons are moving as well as which they
simultaneously help to create. And it then enables one to calculate various
atomic characteristics needed (atomic energy levels, transition
probabilities etc.) or some nuclear-atomic quantities  (e.g. internal conversion
coefficients) as well as other quantities  such as, e.g., Fermi function
required in the \bet-decay studies. That is why the correctness of the atomic
potential, i.e. its ability to describe the reality as well as possible,
 is extremely important.

From the first decade of the last century, when Rutherford introduced the first
realistic atomic model \cite{Ruth11}, it is clear that the atomic potential
must be composed from a Coulombic one, $V_C\sim 1/r$, and an `electronic' part
originated from the electronic cloud. However, as stated above, to calculate
this electronic part (the `screening') is a difficult task.
One way to manage the screening is to use
the so called screening constants, i.e. to replace the atomic number $Z$ by an
``effective'' $Z_{eff}=Z-\sigma$ where $\sigma$ is calculated under various
assumption and tabulated -- see e.g. \cite{Cle,Tho}. This approach, however,
does not supply a uniform potential for atom since the screening constants
are not characteristics of the atom but of the particular atomic shells.

In the twenties of the last century, the statistical Thomas-Fermi
model (potential) \cite{TF27} and its relativistic improvement, Thomas-Fermi-Dirac
model \cite{Dir30}, were suggested. Approximately at the same time the Hartree-Fock
model \cite{HF} was developed. This  approach does not use the conception
of a potential and contains non-local terms. Finally the Hartree-Fock-Slater
(HFS) model \cite{Sla51} where those non-local terms are replaced by an
 averaged potential was created. This one became the most popular approach and
 the HFS potential was often evaluated and tabulated for broad range
 of atomic numbers.

 Another approach is the so called ROEP (relativistic, optimized, effective
 potential) approximation (see, e.g. \cite{Sha53,Tal76}) and its generalization
 RNPOEP (relativistic, numerically parametrized, optimized, effective
 potential) \cite{EffPot}.

In this work we would like to show that the very recent tables \cite{Maldo} of
the  potential based on the RNPOEP method,
when used to calculate some quantities, give surprisingly worse results than
the more than 40 years old HFS tables \cite{Lu}.
%
\sct{Comparison}
The potential \cite{Lu} is obtained by self-consistent solution of the
Dirac equation \cite{Grant61} and the detailed description of the method
utilized is given in \cite{Tuck}. Moreover the effect of finite nucleus
described by the Fermi distribution of the nuclear charge is taken into
account. The potential is tabulated for all atoms (with $Z$=2 to 126)
in an exponentially equidistant mesh.

In the tables \cite{Maldo}, the  finite nucleus is described by a
homogeneously charged
sphere. The resulting potential, however, is not tabulated but is given
in terms of Yukawian functions times a power of $r$. In particular,
$V(r)=-\frac{Z}{2R}(3-\frac{r^2}{R^2})$ for $r<R$, $V(r)=-\frac{1}{r}
[Z-N+1+(N-1)f(r)]$ for $r\geq R$. Here, $Z$ is the atomic number, $N$
is the number of electrons in the atom, $R$ is the atomic radius and
 $f(r)=\sum_{k=1}^{n_C}
{c_k r^{n_k} e^{-\beta_k r}}$. Usually, $n_C$=6 and the parameters $c_k,
n_k$, and $\beta_k$ are results of the calculations and are tabulated
in \cite{Maldo}.

\begin{figure}[h]
\centerline{\includegraphics[clip=on,angle=0,width=16cm]{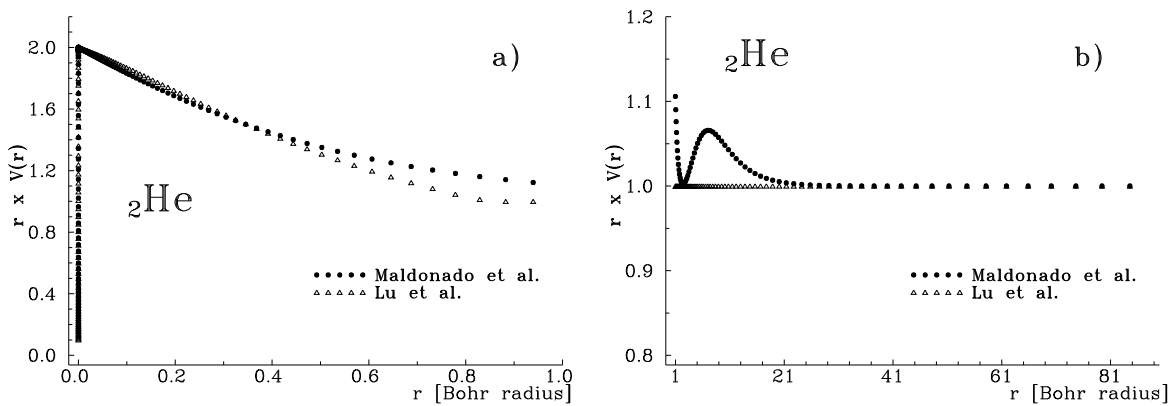}}
\caption{Comparison of the shapes of the potentials of Refs. \cite{Maldo,Lu}
    for neutral atom  \dn{2}He.} \label{f:shapes}
\end{figure}

When we draw the shapes of the two potentials \cite{Maldo,Lu} for the
lightest (except hydrogen) element \dn{2}He, which is presented in
Fig.\ref{f:shapes}, we see a strange behaviour of the potential \cite{Maldo}.
For higher $r$, the atomic potential for the neutral atom should behave as
$\sim 1/r$. The potential \cite{Maldo}, however, shows an unexpected ``wave''
for $r$ between 1 and \asi 20 Bohr radii. The reason of this behaviour is not
clear. One possible reason may be the fact, that the potential \cite{Maldo}
is a combination of analytical function. The effect may be similar to known oscillation of a function {\em interpolated} by polynomials when the
interpolation is done improperly.

The above fact is only indirect and too weak argument against the potential
\cite{Maldo}. To decide which of the two potentials is better, we do two
studies. First, we evaluate the electron binding energies for several
atoms with various $Z$. Comparing them with experimental values indicates
the quality of the particular potentials.
Second, we evaluate some internal conversion coefficients (ICC)
for several transitions
where the experimental values are known. And again, we may compare  the
calculated and experimental values and decide.
%
\subsct{Binding energies}
We have calculated electron energy eigenvalues using the both
potentials \cite{Maldo,Lu}, respectively. To this aim the proper
subroutine of the internal conversion coefficients evaluating program
\cite{Ry} was used. The calculations were performed for a broad
range of atomic numbers, in particular for \dn{17}Cl, \dn{29}Cu,
\dn{36}Kr, \dn{37}Rb, \dn{43}Tc, \dn{62}Sm, \dn{74}W, \dn{76}Os, and
 \dn{88}Ra.
These eigenvalues were compared with the experimental electron binding
energies \cite{Fire}. As an example, the results for the \dn{29}Cu and
\dn{88}Ra are given in Tables~\ref{t:Z29} and \ref{t:Z88}, respectively.

The 8 elements studied contain altogether 136 atomic subshells, this
means that we have 136 energy eigenvalues. From these, in 79 cases the
values calculated with the potential \cite{Lu} are closer to the experimental
binding energies than those calculated with \cite{Maldo}. This is about 58\%
of cases. However, all the cases of better agreement are those on the inner
atomic shells as seen from Table~\ref{t:shells}. This is important since
the binding energies of the outermost electrons are affected by chemical
environment of the atom. Our calculations assume free isolated atom
which need not be (and probably is not) true for the experimental
values \cite{Fire}. Therefore the disagreement of energies at the
outer atomic subshells is not too significant.

As a quantitative measure of the agreement we have evaluated the quantity
\beq
\label{e:delta}
\Delta^2 = \frac{1}{n}\sum_{i=1}^n{(\frac{theor-exp}{exp}\times 100)^2},
\eeq
where $theor$ and $exp$ are theoretical and experimental, respectively,
binding energies and $i$ runs over the atomic subshells taken into
consideration. (The scaling factor 100 has no physical meaning; it is there
to ensure that the resulting $\Delta^2$'s are in a ``readable'' range.)
 Following the above argumentations we took the inner subshells
only which ale enumerated in Table~\ref{t:shells}. Note that the expression
$\Delta^2$ in (\ref{e:delta}) formally resembles the known quantity
$\chi^2$ but it is not -- it does not bear the statistical contents of the
$\chi^2$. Nevertheless, it may be a measure of agreement -- the larger
$\Delta^2$, the poorer agreement. The values of $\Delta^2$ are presented
in Table~\ref{t:shells}, too.
\subsct{Conversion coefficients}
For the calculations of the internal conversion coefficients,
the good description of the atom -- i.e. a good atomic potential --
is needed, too. Therefore, we have evaluated some ICC (for which
experimental data are available) using, respectively, both
atomic potentials \cite{Maldo,Lu}. The comparison with the
mentioned experimental values may supply further arguments which
of the two potentials is ``better''.

For this aim we have chosen the ICC of the transitions of pure
E2 multipolarity, in particular of the transitions 2\up{+}
\ria\ 0\up{+}. These were chosen since their multipolarity is
unambiguously determined and the comparison may not
be disturbed by an effect of multipolarity mixing.

The following data were taken into account (the values of the
transition energies are from \cite{Fire}):
\begin{description}
\item{\bf\elem{62}{152}{Sm}, transition 121.78~keV}  \\
   K/L = 1.74 \plm\ 0.0.06; K/L\dn{3} = 4.22 \plm\ 0.17 \cite{BogZP}\\
   M/L = 0.250 \plm\ 0.015 \cite{Bog106}\\
\item{\bf\elem{66}{162}{Dy}, transition 86.788~keV}  \\
  K/L = 0.62 \plm\ 0.03;  K/L\dn{3} = 1.28 \plm\ 0.06 \cite{BogZP}\\
  M/L = 0.242 \plm\ 0.015 \cite{Bog106}\\
  L\dn{1}/L\dn{2} = 0.136 \plm\ 0.004; L\dn{1}/L\dn{3} = 0.130 \plm\ 0.003;
  L\dn{2}/L\dn{3} = 0.962 \plm\ 0.014 weighted mean from \cite{Ham66,Kar66,Gel66,Erm66}\\
\item{\bf\elem{68}{166}{Er}, transition 80.557~keV}\\
    see Table~\ref{t:erbium}\\
\item{\bf\elem{74}{182}{W}, transition 100.107~keV} \\
  K/L = 0.375 \plm\ 0.027  w.m. from \cite{BogZP,Nils68}\\
  K/L\dn{3} = 0.850 \plm\ 0.026 \cite{BogZP}\\
  M/L = 0.244 \plm\ 0.015 \cite{Bog106}\\
  The next K to O are normalized to L\dn{3}=10000: \cite{Nils67}\\
\indent  K = 7752 \plm\ 444; L = 21789 \plm\ 203; M = 5320 \plm\ 160;
  N = 1138 \plm\ 62; O = 199 \plm\ \ 30\\
  L\dn{1}/L\dn{2} = 0.091 \plm\ 0.015; L\dn{1}/L\dn{3} = 0.099 \plm\ 0.017; L\dn{2}/L\dn{3} = 1.087 \plm\ 0.014; w.m. from \cite{Nils67,Ham66} and references therein\\
  M\dn{1}/M\dn{2} = 0.080 \plm\ 0.030; M\dn{1}/M\dn{3} = 0.082 \plm\ 0.038;
  M\dn{2}/M\dn{3} = 1.031 \plm\ 0.031 w.m. from \cite{Nils67,Hoeg68}\\
  L\dn{3}/M\dn{3} = 3.72 \plm\ 0.07; M/NO = 3.59 \plm\ 0.09 \cite{Nils68}\\
\item{\bf\elem{76}{188}{Os}, transition 155.021~keV}\\
  K/L = 0.840 \plm\ 0.025; K/L\dn{3} = 2.24 \plm\ 0.07 \cite{BogZP}\\
  M/L = 0.255 \plm\ 0.015 \cite{Bog106}\\
  L\dn{1}/L\dn{3} = 0.266 \plm\ 0.005; L\dn{2}/L\dn{3} = 1.38 \plm\ 0.02 \cite{Erm66}\\
\item{\bf\elem{80}{198}{Hg}, transition 411.805~keV}\\
  K/L = 2.69 \plm \ 0.06 w.m. from \cite{BogZP,Kel59}\\
  K/L\dn{3} = 15.15 \plm\ 0.60 \cite{BogZP}\\
  L:M:N:O = 1 : (0.252 \plm\ 0.004) : (0.077 \plm\ 0.004) : (0.018 \plm\ 0.002) \cite{Kel59}\\
  L\dn{1}/L\dn{2} = 0.969 \plm\ 0.027; L\dn{1}/L\dn{3} = 2.221 \plm\ 0.035; L\dn{2}/L\dn{3} = 2.294 \plm\ 0.067 w.m. from \cite{Ham66,Kel59}\\
  K = 0.0308 \plm\ 0.0009 \cite{Pet65}\\
\item{\bf\elem{80}{199}{Hg}, transition 158.371~keV}\\
  All the next values are normalized to L\dn{3}=0.172 \cite{Dra77}:\\
\indent K = 0.284 \plm\ 0.011; L\dn{1} = 0.0387 \plm\ 0.0008; L\dn{2} = 0.251 \plm\ 0.003;\\
\indent M\dn{1} = 0.00943 \plm\ 0.00058; M\dn{2} = 0.0645 \plm\ 0.0034; M\dn{3} = 0.0457 \plm\ 0.0024;\\
\indent M\dn{45} = 0.00105 \plm\ 0.00010; N\dn{1} = 0.00236 \plm\ 0.00017; N\dn{2} = 0.0169 \plm\ 0.0010;\\
\indent  N\dn{3} = 0.0111 \plm\ 0.0006; OPN\dn{67} = .0051 \plm\ 0.0003; \\ \indent Total ICC = 0.903 \plm\ 0.013.
\end{description}
Altogether 64 experimental data items were used.

To compare the agreement with experiment we chose the quantity similar to
that in (\ref{e:delta}), in particular
\beq
\label{e:ICC}
r = (\frac{theor-exp}{\sigma_{exp}})^2
\eeq
(where $\sigma_{exp}$ is the experimental uncertainty), for the individual
ICC or ICC ratios. Note that the $r$ is, in fact, the square of the
residual used in the quantity $\chi^2$. Their sum, however, cannot
be declared to be $\chi^2$ since the particular data items are {\em not}
independent.
Nevertheless the claim ``the higher $r$, the worse agreement'' is
unquestionably true.

In Table~\ref{t:erbium}, there are the experimental and theoretical data for
the E2 transition 80.557~keV in \elem{68}{166}{Er} together with the
resulting values of $r$. We see that the results with the potential \cite{Lu}
are better than those with \cite{Maldo} in 11 cases from 12. And the
agreement in the remaining case (i.e. K/L) is almost the same.

The Table~\ref{t:totagree} shows the overview of the ICC comparison. For
every transition studied, it includes the number of data items at disposal.
In the last three columns are, respectively, the number of cases in favor
(results \cite{Lu} agree better), of cases `neutral' (the agreement is
practically the same), and `contra' (results \cite{Maldo} agree better).
We see that the absolute majority of cases (40 from 64) is `in favor'.
In 16 cases (from 64), both sets of data agree with experiment similarly.
And only in 8 cases the result with potential \cite{Maldo} agrees better
than that with \cite{Lu}.
\sct{Conclusions}
We have studied the quality of  two atomic potentials. One of them,
\cite{Lu}, was published more than 40 years ago (in tabular form), the other,
\cite{Maldo}, (given in form of several coefficients in analytical
formulae) is
quite new. We have checked their quality by the comparison of the
experimental data available with the theoretical results obtained by
use of these two potentials. The data used were of two different areas --
the electron binding energies and the internal conversion coefficients,
respectively. In both these areas it turned out that -- surprisingly --
the older potential \cite{Lu} describes the reality in general better
than does the newest one \cite{Maldo}.

\vspace{3ex}
\noindent{\bf Acknowledgement}  \\ \ \\
This work was supported by the Grant Agency of the Czech
   Republic under contract No. P203/12/1896.
%
%

%
\begin{table}[p]
\bec
\caption{Comparison of binding energies for \dn{29}Cu.}
\ \\
\label{t:Z29}
\begin{tabular}{|l|lll|ll|}
\hline
\multicolumn{1}{|c|}{} & \multicolumn{3}{c|}{energies [eV]} & \multicolumn{2}{c|}{rel. diff. [\%]\up{*)}} \\
\cline{2-6}
\multicolumn{1}{|c|}{shell} & \multicolumn{1}{c}{\cite{Lu}} & \multicolumn{1}{c}{\cite{Maldo}} & \multicolumn{1}{c|}{exp. \cite{Fire}} & \multicolumn{1}{c}{\cite{Lu}} & \multicolumn{1}{c|}{\cite{Maldo}} \\
\hline
\multicolumn{1}{|c|}{K} & \multicolumn{1}{r}{8945.79} & \multicolumn{1}{r}{8853.25} & \multicolumn{1}{r|}{8978.9} & \multicolumn{1}{r}{-0.37} & \multicolumn{1}{r|}{-1.40} \\
\multicolumn{1}{|c|}{L\dn{1}} & \multicolumn{1}{r}{1086.66} & \multicolumn{1}{r}{1064.66} & \multicolumn{1}{r|}{1096.1} & \multicolumn{1}{r}{-0.86} & \multicolumn{1}{r|}{-2.87} \\
\multicolumn{1}{|c|}{L\dn{2}} & \multicolumn{1}{r}{959.05} & \multicolumn{1}{r}{931.67} & \multicolumn{1}{r|}{951.0} & \multicolumn{1}{r}{0.85} & \multicolumn{1}{r|}{-2.03} \\
\multicolumn{1}{|c|}{L\dn{3}} & \multicolumn{1}{r}{937.89} & \multicolumn{1}{r}{911.20} & \multicolumn{1}{r|}{931.1} & \multicolumn{1}{r}{0.73} & \multicolumn{1}{r|}{-2.14} \\
\multicolumn{1}{|c|}{M\dn{1}} & \multicolumn{1}{r}{121.41} & \multicolumn{1}{r}{119.25} & \multicolumn{1}{r|}{119.8} & \multicolumn{1}{r}{1.34} & \multicolumn{1}{r|}{-0.46} \\
\multicolumn{1}{|c|}{M\dn{2}} & \multicolumn{1}{r}{80.63} & \multicolumn{1}{r}{78.03} & \multicolumn{1}{r|}{73.6} & \multicolumn{1}{r}{9.55} & \multicolumn{1}{r|}{6.02} \\
\multicolumn{1}{|c|}{M\dn{3}} & \multicolumn{1}{r}{77.94} & \multicolumn{1}{r}{75.52} & \multicolumn{1}{r|}{73.6} & \multicolumn{1}{r}{5.90} & \multicolumn{1}{r|}{2.61} \\
\multicolumn{1}{|c|}{M\dn{4}} & \multicolumn{1}{r}{10.15} & \multicolumn{1}{r}{8.58} & \multicolumn{1}{r|}{1.6} & \multicolumn{1}{r}{534.37} & \multicolumn{1}{r|}{436.25} \\
\multicolumn{1}{|c|}{M\dn{5}} & \multicolumn{1}{r}{9.84} & \multicolumn{1}{r}{8.31} & \multicolumn{1}{r|}{1.6} & \multicolumn{1}{r}{515.00} & \multicolumn{1}{r|}{419.38} \\
\hline
\end{tabular}\\
\bfl
\up{*)} rel. diff. = $\frac{{\rm [x] - exp}}{{\rm exp}} \times 100$
\efl
\eec
\end{table}
\begin{table}[p]
\bec
\caption{Comparison of binding energies for \dn{88}Ra.}
\ \\
\label{t:Z88}
\begin{tabular}{|l|lll|ll|}
\hline
\multicolumn{1}{|c|}{} & \multicolumn{3}{c|}{energies [eV]} & \multicolumn{2}{c|}{rel. diff. [\%]\up{*)}} \\
\cline{2-6}
\multicolumn{1}{|c|}{shell} & \multicolumn{1}{c}{\cite{Lu}} & \multicolumn{1}{c}{\cite{Maldo}} & \multicolumn{1}{c|}{exp. \cite{Fire}} & \multicolumn{1}{c}{\cite{Lu}} & \multicolumn{1}{c|}{\cite{Maldo}} \\
\hline
\multicolumn{1}{|c|}{K} & \multicolumn{1}{r}{104470.20} & \multicolumn{1}{r}{103862.93} & \multicolumn{1}{r|}{103915} & \multicolumn{1}{r}{0.53} & \multicolumn{1}{r|}{-0.05} \\
\multicolumn{1}{|c|}{L\dn{1}} & \multicolumn{1}{r}{19255.93} & \multicolumn{1}{r}{19112.69} & \multicolumn{1}{r|}{19232} & \multicolumn{1}{r}{0.12} & \multicolumn{1}{r|}{-0.62} \\
\multicolumn{1}{|c|}{L\dn{2}} & \multicolumn{1}{r}{18571.80} & \multicolumn{1}{r}{18408.23} & \multicolumn{1}{r|}{18484} & \multicolumn{1}{r}{0.48} & \multicolumn{1}{r|}{-0.41} \\
\multicolumn{1}{|c|}{L\dn{3}} & \multicolumn{1}{r}{15460.97} & \multicolumn{1}{r}{15337.17} & \multicolumn{1}{r|}{15444} & \multicolumn{1}{r}{0.11} & \multicolumn{1}{r|}{-0.69} \\
\multicolumn{1}{|c|}{M\dn{1}} & \multicolumn{1}{r}{4801.88} & \multicolumn{1}{r}{4736.89} & \multicolumn{1}{r|}{4822} & \multicolumn{1}{r}{-0.42} & \multicolumn{1}{r|}{-1.77} \\
\multicolumn{1}{|c|}{M\dn{2}} & \multicolumn{1}{r}{4486.08} & \multicolumn{1}{r}{4411.20} & \multicolumn{1}{r|}{4483} & \multicolumn{1}{r}{0.07} & \multicolumn{1}{r|}{-1.60} \\
\multicolumn{1}{|c|}{M\dn{3}} & \multicolumn{1}{r}{3775.30} & \multicolumn{1}{r}{3716.60} & \multicolumn{1}{r|}{3785} & \multicolumn{1}{r}{-0.26} & \multicolumn{1}{r|}{-1.81} \\
\multicolumn{1}{|c|}{M\dn{4}} & \multicolumn{1}{r}{3260.29} & \multicolumn{1}{r}{3194.91} & \multicolumn{1}{r|}{3248} & \multicolumn{1}{r}{0.38} & \multicolumn{1}{r|}{-1.63} \\
\multicolumn{1}{|c|}{M\dn{5}} & \multicolumn{1}{r}{3112.27} & \multicolumn{1}{r}{3049.39} & \multicolumn{1}{r|}{3105} & \multicolumn{1}{r}{0.23} & \multicolumn{1}{r|}{-1.79} \\
\multicolumn{1}{|c|}{N\dn{1}} & \multicolumn{1}{r}{1192.01} & \multicolumn{1}{r}{1163.60} & \multicolumn{1}{r|}{1208} & \multicolumn{1}{r}{-1.32} & \multicolumn{1}{r|}{-3.68} \\
\multicolumn{1}{|c|}{N\dn{2}} & \multicolumn{1}{r}{1050.11} & \multicolumn{1}{r}{1020.05} & \multicolumn{1}{r|}{1055} & \multicolumn{1}{r}{-0.46} & \multicolumn{1}{r|}{-3.31} \\
\multicolumn{1}{|c|}{N\dn{3}} & \multicolumn{1}{r}{867.94} & \multicolumn{1}{r}{842.15} & \multicolumn{1}{r|}{879} & \multicolumn{1}{r}{-1.26} & \multicolumn{1}{r|}{-4.19} \\
\multicolumn{1}{|c|}{N\dn{4}} & \multicolumn{1}{r}{637.33} & \multicolumn{1}{r}{612.57} & \multicolumn{1}{r|}{636} & \multicolumn{1}{r}{0.21} & \multicolumn{1}{r|}{-3.68} \\
\multicolumn{1}{|c|}{N\dn{5}} & \multicolumn{1}{r}{603.24} & \multicolumn{1}{r}{579.10} & \multicolumn{1}{r|}{603} & \multicolumn{1}{r}{0.04} & \multicolumn{1}{r|}{-3.96} \\
\multicolumn{1}{|c|}{N\dn{6}} & \multicolumn{1}{r}{299.86} & \multicolumn{1}{r}{278.77} & \multicolumn{1}{r|}{287} & \multicolumn{1}{r}{4.48} & \multicolumn{1}{r|}{-2.87} \\
\multicolumn{1}{|c|}{N\dn{7}} & \multicolumn{1}{r}{291.25} & \multicolumn{1}{r}{270.51} & \multicolumn{1}{r|}{279} & \multicolumn{1}{r}{4.39} & \multicolumn{1}{r|}{-3.04} \\
\multicolumn{1}{|c|}{O\dn{1}} & \multicolumn{1}{r}{255.99} & \multicolumn{1}{r}{242.95} & \multicolumn{1}{r|}{251} & \multicolumn{1}{r}{1.99} & \multicolumn{1}{r|}{-3.21} \\
\multicolumn{1}{|c|}{O\dn{2}} & \multicolumn{1}{r}{202.81} & \multicolumn{1}{r}{189.64} & \multicolumn{1}{r|}{197} & \multicolumn{1}{r}{2.95} & \multicolumn{1}{r|}{-3.74} \\
\multicolumn{1}{|c|}{O\dn{3}} & \multicolumn{1}{r}{162.87} & \multicolumn{1}{r}{150.54} & \multicolumn{1}{r|}{153} & \multicolumn{1}{r}{6.45} & \multicolumn{1}{r|}{-1.61} \\
\multicolumn{1}{|c|}{O\dn{4}} & \multicolumn{1}{r}{84.91} & \multicolumn{1}{r}{73.51} & \multicolumn{1}{r|}{72} & \multicolumn{1}{r}{17.93} & \multicolumn{1}{r|}{2.10} \\
\multicolumn{1}{|c|}{O\dn{5}} & \multicolumn{1}{r}{79.13} & \multicolumn{1}{r}{67.88} & \multicolumn{1}{r|}{66} & \multicolumn{1}{r}{19.89} & \multicolumn{1}{r|}{2.85} \\
\multicolumn{1}{|c|}{P\dn{1}} & \multicolumn{1}{r}{41.57} & \multicolumn{1}{r}{35.99} & \multicolumn{1}{r|}{31} & \multicolumn{1}{r}{34.10} & \multicolumn{1}{r|}{16.10} \\
\multicolumn{1}{|c|}{P\dn{2}} & \multicolumn{1}{r}{26.19} & \multicolumn{1}{r}{22.19} & \multicolumn{1}{r|}{20} & \multicolumn{1}{r}{30.95} & \multicolumn{1}{r|}{10.95} \\
\multicolumn{1}{|c|}{P\dn{3}} & \multicolumn{1}{r}{19.32} & \multicolumn{1}{r}{16.83} & \multicolumn{1}{r|}{12} & \multicolumn{1}{r}{61.00} & \multicolumn{1}{r|}{40.25} \\
\hline
\end{tabular}\\
\bfl
\up{*)} rel. diff. = $\frac{{\rm [x] - exp}}{{\rm exp}} \times 100$
\efl
\eec
\end{table}
\begin{table}[p]
\bec
\caption{List of atomic subshells where eigenenergies with potential \cite{Lu}
     agree better with experiment.}
\ \\
\label{t:shells}
\begin{tabular}{|l|l|l|l|l|l|}
\hline
 & \multicolumn{2}{c|}{subshells} & \multicolumn{3}{c|}{$\Delta^2$ \up{**)}} \\
\hline
\multicolumn{1}{|c|}{element} & \multicolumn{1}{c|}{last} & \multicolumn{1}{c|}{in favor\up{*)}} & \multicolumn{1}{c|}{$n$} & \multicolumn{1}{c|}{\cite{Lu}} & \multicolumn{1}{c|}{\cite{Maldo}} \\
\hline
\multicolumn{1}{|c|}{\dn{17}Cl} & \multicolumn{1}{c|}{M\dn{3}} & K,L\dn{1} & \multicolumn{1}{r|}{2} & \multicolumn{1}{r|}{0.45} & \multicolumn{1}{r|}{9.10} \\
\multicolumn{1}{|c|}{\dn{29}Cu} & \multicolumn{1}{c|}{M\dn{5}} & K,L\dn{1-3},M\dn{1} & \multicolumn{1}{r|}{4} & \multicolumn{1}{r|}{0.53} & \multicolumn{1}{r|}{4.72} \\
\multicolumn{1}{|c|}{\dn{36}Kr} & \multicolumn{1}{c|}{N\dn{3}} & K,L\dn{1-3},M\dn{1-5},N\dn{1} & \multicolumn{1}{r|}{10} & \multicolumn{1}{r|}{2.67} & \multicolumn{1}{r|}{20.92} \\
\multicolumn{1}{|c|}{\dn{37}Rb} & \multicolumn{1}{c|}{O\dn{1}} & K,L\dn{1-3},M\dn{1-3} & \multicolumn{1}{r|}{7} & \multicolumn{1}{r|}{1.17} & \multicolumn{1}{r|}{21.39} \\
\multicolumn{1}{|c|}{\dn{43}Tc} & \multicolumn{1}{c|}{N\dn{5}} & K,L\dn{1-3},M\dn{1-3} & \multicolumn{1}{r|}{7} & \multicolumn{1}{r|}{0.39} & \multicolumn{1}{r|}{5.13} \\
\multicolumn{1}{|c|}{\dn{62}Sm} & \multicolumn{1}{c|}{N\dn{6}} & K,L\dn{1-3},M\dn{1-5},N\dn{1,3} & \multicolumn{1}{r|}{11} & \multicolumn{1}{r|}{0.47} & \multicolumn{1}{r|}{4.61} \\
\multicolumn{1}{|c|}{\dn{74}W} & \multicolumn{1}{c|}{O\dn{4}} & L\dn{1-3},M\dn{1-5},N\dn{1-5} & \multicolumn{1}{r|}{13} &
\multicolumn{1}{r|}{0.31} & \multicolumn{1}{r|}{7.63} \\
\multicolumn{1}{|c|}{\dn{76}Os} & \multicolumn{1}{c|}{O\dn{3}} & K,L\dn{1-3},M\dn{1-5},N\dn{1,3} & \multicolumn{1}{r|}{12} & \multicolumn{1}{r|}{0.19} & \multicolumn{1}{r|}{3.69} \\
\multicolumn{1}{|c|}{\dn{88}Ra} & \multicolumn{1}{c|}{P\dn{3}} & K,L\dn{1-3},M\dn{1-5},N\dn{1-5},O\dn{1,2} & \multicolumn{1}{r|}{16} & \multicolumn{1}{r|}{1.08} & \multicolumn{1}{r|}{6.96} \\
\hline
\end{tabular}\\
\bfl
\up{*)} the subshells where the results with \cite{Lu} are
  closer to experiment \\
\up{**)} see Eq.(\ref{e:delta})
\efl
\eec
\end{table}
\begin{table}[p]
\bec
\caption{Comparison of ICC ratios for \elem{68}{166}{Er}.}
\ \\
\label{t:erbium}
\begin{tabular}{|l|r@{.}l@{ }ll|r@{.}lr@{.}l|r@{.}lr@{.}l|}
\hline
 & \multicolumn{3}{c}{} &  & \multicolumn{4}{c|}{theory} & \multicolumn{4}{c|}{r\up{2}\ \up{*)}} \\
\cline{6-13}
shells & \multicolumn{3}{c}{experiment} & \multicolumn{1}{c|}{ref.} & \multicolumn{2}{c}{\cite{Lu}} & \multicolumn{2}{c|}{\cite{Maldo}} & \multicolumn{2}{c}{\cite{Lu}} & \multicolumn{2}{c|}{\cite{Maldo}} \\
\hline
K/L & 0&416 & \plm\  0.028 & \cite{BogZP,Nils68} & 0&4121 & 0&4123 & 0&019 & 0&017 \\
K/L\dn{3} & 0&805 & \plm\  0.034 & \cite{BogZP} & 0&8466 & 0&8484 & 1&497 & 1&629 \\
L\dn{3}/M\dn{3} & 4&00 & \plm\  0.05 & \cite{Nils68} & 4&0748 & 4&0831 & 2&238 & 2&762 \\
L\dn{1}/L\dn{3} & 0&0864 & \plm\  0.0011 & \cite{Kar66,Gel66,Erm66} & 0&0812 & 0&0805 & 22&347 & 28&769 \\
L\dn{2}/L\dn{3} & 0&958 & \plm\  0.010 & \cite{Kar66,Gel66,Erm66} & 0&9730 & 0&9771 & 2&250 & 3&648 \\
L\dn{1}/L\dn{2} & 0&0910 & \plm\  0.0037 & \cite{Kar66} & 0&0835 & 0&0824 & 4&109 & 5&402 \\
M/L & 0&250 & \plm\  0.018 & \cite{Bog106} & 0&2440 & 0&2432 & 0&006 & 0&143 \\
M/NO & 3&78 & \plm\  0.09 & \cite{Nils68} & 3&798\  & 3&861\  & 0&040 & 0&810 \\
M\dn{1}\up{**)} & 7&90 & \plm\  0.18 & \cite{Hoeg68} & 7&500\  & 7&462\  & 4&938 & 5&921 \\
M\dn{2}\up{**)} & 93&4 & \plm\  0.8 & \cite{Hoeg68} & 94&75\ \  & 94&96\ \  & 2&848 & 3&803 \\
M\dn{4}\up{**)} & 1&05 & \plm\  0.04 & \cite{Hoeg68} & 1&049\  & 1&007\  & 0&001 & 1&156 \\
M\dn{5}\up{**)} & 1&05 & \plm\  0.05 & \cite{Hoeg68} & 0&974\  & 0&936\  & 2&310 & 5&198 \\
\hline
\end{tabular}\\
\bfl
\up{*)} see Eq.(\ref{e:ICC})\\
\up{**)} these subshells are normalized to M\dn{3}=100\\
\efl
\eec
\end{table}
\begin{table}[p]
\bec
\caption{Overview of the agreement with experimental data for the two
sets of the ICC -- with potentials \cite{Lu} and \cite{Maldo},
respectively.}
\ \\
\label{t:totagree}
\begin{tabular}{|l|c|ccc|}
\hline
element & \multicolumn{1}{r|}{no. of data} & \multicolumn{1}{r}{pro\up{*)}} & \multicolumn{1}{r}{neu\up{**)}} & \multicolumn{1}{r|}{con\up{***)}} \\
\hline
\elem{62}{152}{Sm} & 
   3 & 2 & 1 & 0 \\
\elem{66}{162}{Dy} & 
   6 & 2 & 2 & 2\\
\elem{68}{166}{Er} & 
   12 & 11 & 1 & 0\\
\elem{74}{182}{W} & 
   16 & 7 & 7 & 2\\
\elem{76}{188}{Os} & 
  5 &  4 &  0  &  1\\
\elem{80}{198}{Hg} & 
    9  &  6 &  2  &  1\\
\elem{80}{199}{Hg} & 
    13 &  8  &  3  &  2\\
\hline
total & 
   64 &  40  &  16  &  8\\
\hline
\end{tabular}\\
\bfl
\up{*)} results `in favor', for \cite{Lu} are better\\
\up{**)} agreement comparable for both \cite{Lu} and \cite{Maldo}\\
\up{***)} results for \cite{Maldo} are better
\efl
\eec
\end{table}
\end{document}